\documentclass[5p,twocolumn,times]{elsarticle}
\journal{Physics Letters B}

\usepackage{graphicx}
\usepackage{amsmath}
\usepackage{amssymb}

\def\ket#1{|#1\rangle}
\def\bra#1{\langle#1|}
\def\bz{\beta_0}

\newcommand{\ba}{\begin{eqnarray}}
\newcommand{\ea}{\end{eqnarray}}
\newcommand{\bmath}{\begin{mathletters}}
\newcommand{\emath}{\end{mathletters}}
\newcommand{\ban}{\begin{eqnarray*}}
\newcommand{\ean}{\end{eqnarray*}}

\newcommand{\bsub}{\begin{subequations}}
\newcommand{\esub}{\end{subequations}}

\begin{document}


\title
{Evolution of order and chaos across a first-order quantum phase transition}

\author{A. Leviatan\corref{cor1}}
\ead{ami@phys.huji.ac.il}

\author{M.~Macek}
\ead{mmacek@phys.huji.ac.il}

\cortext[cor1]{Corresponding author}

\address{Racah Institute of Physics, The Hebrew University, 
Jerusalem 91904, Israel}

\begin{abstract}
We study the evolution of the dynamics 
across a generic first order quantum phase transition in an interacting 
boson model of nuclei. 
The dynamics inside the phase coexistence region exhibits 
a very simple pattern. 
A~classical analysis reveals a robustly regular dynamics confined to 
the deformed region and well separated from a chaotic dynamics 
ascribed to the spherical region. A~quantum analysis discloses regular 
bands of states in the deformed region, which persist to energies well 
above the phase-separating barrier, in the face of a complicated 
environment. The impact of kinetic collective rotational terms on this 
intricate interplay of order and chaos is investigated.
\end{abstract}



\maketitle

Quantum phase transitions (QPTs) are qualitative changes in the properties 
of a physical system induced by a variation of parameters $\lambda$ in the 
quantum Hamiltonian $\hat{H}(\lambda)$~\cite{ref:Hert76,ref:Gilm79}. 
Such ground-state transformations have received considerable attention in 
recent years and have found a variety of applications in many areas of 
physics and chemistry~\cite{carr}. 
The competing interactions in the Hamiltonian that 
drive these transitions, can affect dramatically the nature of the 
dynamics and, in some cases, lead to the emergence of quantum chaos. 
This effect has been observed in quantum optics models of $N$ 
two-level atoms interacting with a 
single-mode radiation field~\cite{ref:Emar03}, where 
the onset of chaos is triggered by continuous QPTs. 
In this case, the underlying 
mean-field (Landau) potential $V(\lambda)$ has a single minimum 
which evolves continuously into another minimum. 
The situation is more complex for discontinuous (first-order) QPTs. 
Here $V(\lambda)$ develops multiple minima 
that coexist in a range of $\lambda$ values 
and cross at the critical point, $\lambda\!=\!\lambda_c$. 
Understanding the nature of the underlying dynamics in such circumstances 
is a primary goal of the present Letter. 

The interest in first-order QPTs stems from their key role in 
phase-coexistence phenomena at zero temperature. 
In condensed matter physics, it has been recently recognized that, 
for clean samples, the nature of the QPT becomes discontinuous as the 
critical-point is approached. 
Examples are offered by the metal-insulator Mott 
transition~\cite{maria04}, itinerant magnets~\cite{Pf05}, 
heavy-fermion superconductors~\cite{Pf09}, 
and quantum Hall bilayers~\cite{karm09}. 
First-order QPTs are relevant to shape-coexistence in 
mesoscopic systems, such as atomic nuclei~\cite{ref:Cejn10}, and to 
optimization problems in quantum computing~\cite{young10}.

Hamiltonians describing first-order QPTs are often non-integrable, 
hence their dynamics is mixed. They form a subclass among the family of 
generic Hamiltonians with a mixed phase space, 
in which regular and chaotic motion coexist. 
In the present Letter, we wish to illuminate, in a transparent manner, 
those aspects of this mixed dynamics 
which reflect the first-order~transition. 
For that purpose, we employ an interacting boson model 
which describes such QPTs between spherical and axially-deformed nuclei. 
Our main results are that, (i) in spite of the abrupt structural changes 
taking place across a first-order QPT, the dynamics in the coexistence 
region exhibits a very simple pattern. A robustly regular dynamics is 
confined to the deformed region, and is well separated from the chaotic 
dynamics ascribed to the spherical region. (ii) The deviations 
from this marked separation is largely due to kinetic collective 
rotational terms in the Hamiltonian.  
This simple pattern of mixed dynamics was initially observed 
at the critical point, $\lambda\!=\!\lambda_c$~\cite{ref:MacLev11}. 
Here we show it to be a hallmark of the whole coexistence region. 
Simply divided phase spaces were encountered 
in billiard systems~\cite{Bunim01,dietz07}, 
which are generated by the free motion of a point particle inside a 
closed domain whose geometry governs the 
amount of chaoticity. Here, in contrast, we consider 
many-body interacting systems undergoing QPTs, 
where the onset of chaos is governed by a change of coupling constants 
in the Hamiltonian. 

The interacting boson model (IBM)~\cite{ref:Iac87} describes quadrupole 
collective states in nuclei in terms of a system of $N$ 
monopole ($s$) and quadrupole ($d$) bosons, representing valence 
nucleon pairs. 
The Hamiltonian conserves the total boson number $N$ and 
angular momentum $L$. 
Its geometric visualization is obtained by a potential surface, 
$V(\beta,\gamma)\!=\! \bra{\beta,\gamma;N}\hat{H}\ket{\beta,\gamma;N}$, 
defined by the expectation value of the 
Hamiltonian in the intrinsic condensate state 
$\vert\beta,\gamma ; N \rangle \!=\! 
(N!)^{-1/2}[\Gamma^{\dagger}_{c}(\beta,\gamma)]^N\vert 0\rangle$,  
where $\Gamma^{\dagger}_{c}(\beta,\gamma) \!=\! 
{\textstyle\frac{1}{\sqrt{2}}}
[\beta\cos\gamma d^{\dagger}_{0} + \beta\sin{\gamma} 
{\textstyle\frac{1}{\sqrt{2}}}( d^{\dagger}_{2} + d^{\dagger}_{-2}) 
+ \sqrt{2-\beta^2}s^{\dagger}]$~\cite{ref:Gino80}. 
Here $(\beta,\gamma)$ are quadrupole shape parameters whose values 
$(\beta_{\mathrm{eq}},\gamma_{\mathrm{eq}})$ at the global minimum 
of $V(\beta,\gamma)$ 
define the equilibrium shape for a given Hamiltonian. QPTs 
between such stable shapes have been studied extensively
in the IBM framework~\cite{ref:Diep80,ref:Cejnar09,ref:Iac11}
and are manifested empirically in nuclei~\cite{ref:Cejn10}. 
Their nature is dictated by the topology of 
$V(\beta,\gamma)$, which serves as a Landau potential. 
For that reason, in studying such QPTs, 
it is convenient to resolve the Hamiltonian into two 
parts~\cite{ref:Lev87}, 
\ba
\hat{H} = \hat{H}_{\mathrm{int}} + \hat{H}_{\mathrm{col}} ~.
\label{eq:H}
\ea
The intrinsic part ($\hat{H}_{\mathrm{int}}$) 
determines the potential surface $V(\beta,\gamma)$, 
while the collective part ($\hat{H}_{\mathrm{col}}$)
is composed of kinetic terms which do not affect the shape of 
$V(\beta,\gamma)$.

Focusing on first-order QPTs between 
stable spherical ($\beta_{\mathrm{eq}}=0$) and prolate-deformed 
($\beta_{\mathrm{eq}}>0$, $\gamma_{\mathrm{eq}}=0$) shapes, 
the intrinsic Hamiltonian reads
\bsub
\label{eq:Hint}
\ba
\hat{H}_\mathrm{int}^{I}(\rho)/\bar{h}_2 &=& 
2(1\!-\! \rho^2\beta_0^{2})\hat{n}_d(\hat{n}_d \!-\! 1)
+\beta_0^2 R^{\dag}_2 \cdot\tilde{R}_2 ~,\qquad
\label{eq:H1} \\
\hat{H}_\mathrm{int}^{II}(\xi)/ \bar{h}_2 &=& 
\xi P^{\dag}_0 P_0 + 
P^{\dag}_2 \cdot \tilde{P}_2 ~,  
\label{eq:H2} 
\ea
\esub
where $\hat{n}_d \!=\! \sum_{\mu}d^{\dag}_{\mu} d_{\mu}$ is 
the $d$-boson number operator, 
$R^{\dag}_{2\mu}(\rho) \!=\! \sqrt{2}s^\dag d^\dag_\mu + 
\rho\sqrt{7}(d^\dag d^\dag)^{(2)}_\mu$, 
$P^{\dag}_0(\beta_0) \!=\! 
d^\dag \cdot d^\dag - \beta_0^2 (s^\dag)^2$ 
and 
$P^{\dag}_{2\mu}(\beta_0) \!=\! 
\sqrt{2}\beta_0 s^\dag d^\dag_\mu + 
\sqrt{7}(d^\dag d^\dag)^{(2)}_\mu$. 
Here
$\tilde{R}_{2\mu} \!=\! (-1)^{\mu}R_{2,-\mu}$, 
$\tilde{P}_{2\mu} \!=\! (-1)^{\mu}P_{2,-\mu}$ 
and the dot implies a scalar product. 
Scaling by $\overline{h}_2\equiv h_2/ N(N-1)$ 
is used throughout, to facilitate the comparison with the classical limit. 
The control parameters that drive the QPT are $\rho$ and $\xi$, 
with $0\leq\rho\leq\beta_{0}^{-1}$ and $\xi\geq 0$, while $\beta_0$ 
is a constant. For the indicated ranges, the intrinsic 
Hamiltonians in the spherical [$\hat{H}_\mathrm{int}^{I}(\rho)$] 
and deformed [$\hat{H}_\mathrm{int}^{II}(\xi)$] phases have 
the intrinsic states 
$\vert\beta_{\mathrm{eq}},\gamma_{\mathrm{eq}} ; N \rangle$ with, 
respectively, $\beta_{\mathrm{eq}}\!=\!0$ and 
[$\beta_{\mathrm{eq}}\!>\!0,\gamma_{\mathrm{eq}}\!=\!0$], 
as zero-energy ground states. For large $N$, the normal modes 
of $\hat{H}_\mathrm{int}^{I}(\rho)$ [$\hat{H}_\mathrm{int}^{II}(\xi)]$ 
involve quadrupole [both $\beta$ and $\gamma$] vibrations 
about the spherical [deformed] global minimum,  
with frequency $\epsilon \!=\! 2\bar{h}_2N\beta_{0}^2$ 
[$\epsilon_\beta \!=\! 2\bar{h}_2 N\bz^2 (2\xi + 1),\,
\epsilon_\gamma \!=\! 18 \bar{h}_2 N \bz^2 (1+\bz^2)^{-1}$].
The two Hamiltonians coincide at the critical point 
$\rho_c\!=\!\bz^{-1}$ and $\xi_c \!=\!0$: 
$\hat{H}_\mathrm{int}^{I}(\rho_c) \!=\! \hat{H}_\mathrm{int}^{II}(\xi_c)$, 
being equal to the Hamiltonian studied in~\cite{ref:MacLev11}.

The classical limit of the IBM is obtained through the use of coherent 
states and taking $N\rightarrow\infty$, with 
$1/N$ playing the role of $\hbar$~\cite{ref:Hatc82}. 
Number conservation ensures that phase space is 10-dimensional 
and can be phrased in terms of two shape (deformation) variables, 
three orientation (Euler) angles and their 
conjugate momenta. The classical Hamiltonian obtained involves 
complicated expressions (including square roots) of these variables.
Setting all momenta to zero, yields the classical potential which 
is identical to $V(\beta,\gamma)$ mentioned above. 
Chaotic properties of the IBM have been 
studied extensively~\cite{ref:Whel93}, albeit, 
with a simplified Hamiltonian, 
giving rise to extremely low barrier and narrow coexistence region. 
The recent identification of IBM Hamiltonians without such 
restrictions~\cite{ref:Lev06} enables, for the first time, 
a comprehensive analysis across a generic first-order QPT. 

\begin{figure}[t]
\begin{center}
\includegraphics[width=0.99\linewidth]{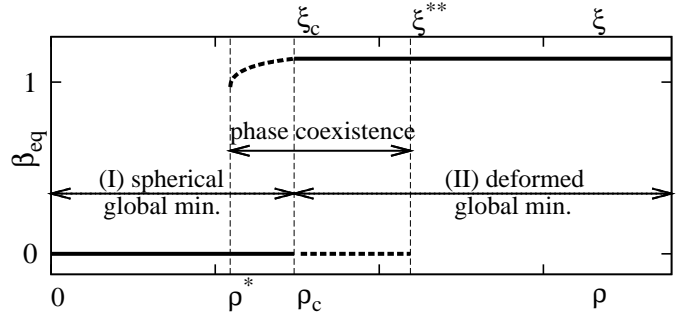}
\end{center}
\caption{Behavior of the order parameter, $\beta_{\mathrm{eq}}$, 
as a function of the control parameters ($\rho,\xi$) of the 
Hamiltonian~(\ref{eq:Hint}). 
Here 
$\rho^{*},\, (\rho_c,\,\xi_c),\, \xi^{**}$, are the spinodal, critical 
and anti-spinodal points, respectively, given by 
$\rho^{*}\!=\!\tfrac{1}{\sqrt{6}}[-(r^2 -4r\!+\!1)+
(r+1)\sqrt{(r+1)(r-1/3)}\,]^{1/2},\, 
(\rho_c\!=\!\bz^{-1},\, \xi_c\!=\!0),\, 
\xi^{**}\!=\!(1+\bz^2)^{-1}$, with $r\!\equiv\!\beta_{0}^{-2}$. 
The deformation at the 
global (local) minimum of the Landau potential (\ref{Vpot}) is marked 
by solid (dashed) lines. 
$\beta_{\mathrm{eq}}\!=\!0$ 
[$\beta_{\mathrm{eq}}\!=\!\sqrt{2}\bz(1+\bz^2)^{-1/2}$] 
on the spherical [deformed] side, with values shown correspond
to $\beta_0 \!=\! 1.35$.} 
\label{fig:1} 
\end{figure}

\begin{figure}[!t]
\begin{center}
\includegraphics[width=1\linewidth]{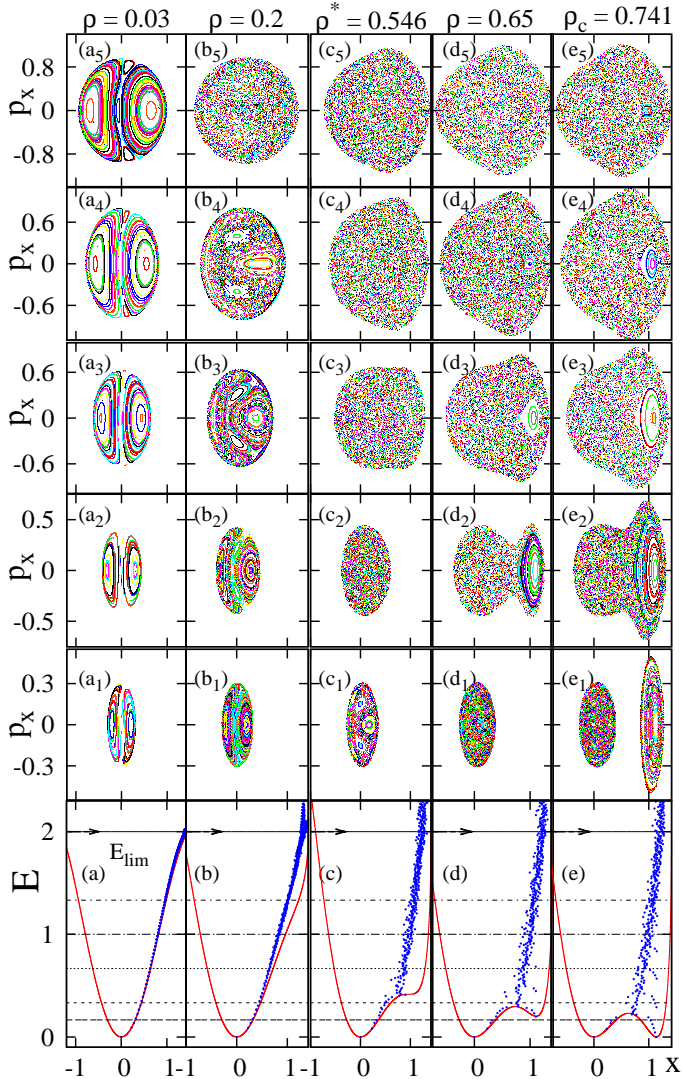}
\end{center}
\caption{(Color online). 
Poincar\'e sections (upper five rows) depicting the classical dynamics 
of $\hat{H}^{I}_{\mathrm{int}}(\rho)$~(\ref{eq:H1})  
with~$h_2\!=\!1,\,\beta_0\!=\!1.35$, for several values of $\rho\leq\rho_c$. 
The bottom row displays the corresponding classical potentials 
$V^{(I)}(\rho)$~(\ref{eq:V1}). 
The five energies, below $E_\mathrm{lim}\!=\!2h_2$, at which the 
sections were calculated consecutively, are indicated by horizontal 
dashed lines.
The Peres lattices
$\{x_i,E_i\}$, portraying the quantum dynamics for eigenstates $\ket{i}$ 
of $\hat{H}^{I}_{\mathrm{int}}(\rho)$ with 
$L\!=\!0$ and $N\!=\!80$, are overlayed on the 
classical potentials $V(x,y=0)$.}
\label{fig:2}
\end{figure}

\begin{figure}[!t]
\begin{center}
\includegraphics[width=0.98\linewidth]{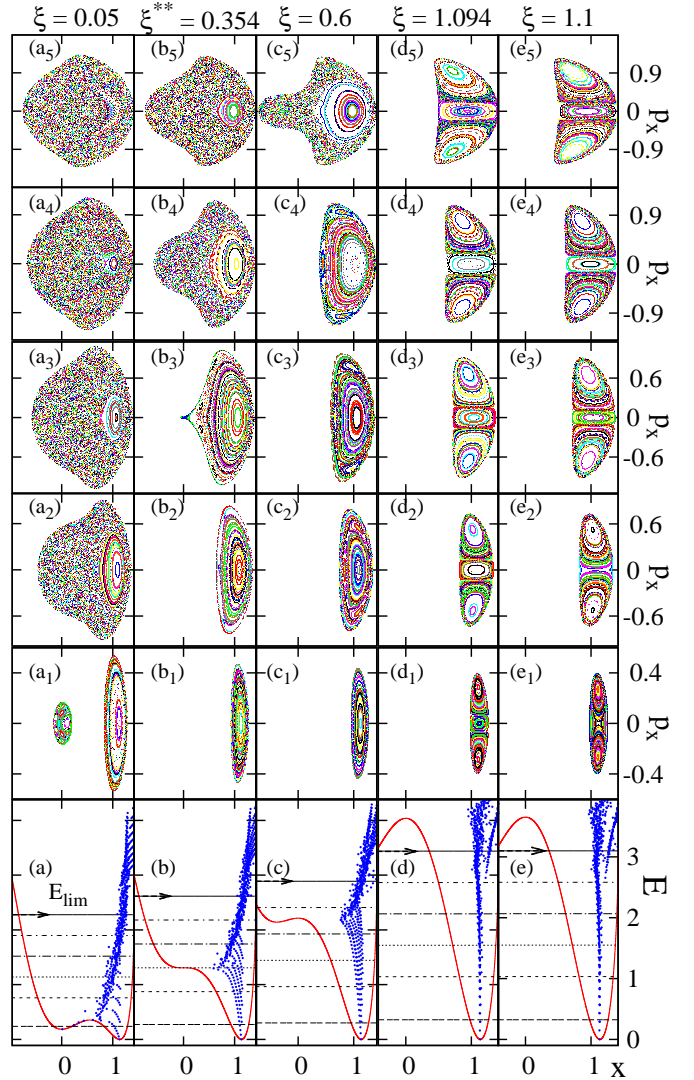}
\end{center}
\caption{(Color online). 
Same as in Fig.~\ref{fig:2} but for the intrinsic Hamiltonian 
$\hat{H}^{II}_{\mathrm{int}}(\xi)$~(\ref{eq:H2}) and classical potential 
$V^{(II)}(\xi)$~(\ref{eq:V2}), 
with $\xi>\xi_c$ and $E_\mathrm{lim}\!=\!h_2(2+\xi)$.
Notice in the Peres lattices $\{x_i,E_i\}$ at the bottom row, 
the sequences of regular states in the deformed region $x\approx 1$ 
[also observed in Figs.~2(d)-2(e)]. 
The lowest sequence consists of 
$(N\!=\!80,L\!=\!0)$ bandhead states of 
the ground $g(K=0)$ and $\beta^n(K=0)$ bands. 
Regular sequences at higher energy correspond to 
$\beta^n\gamma^2(K=0)$, $\beta^n\gamma^{4}(K=0)$ 
bands, etc.} 
\label{fig:3} 
\end{figure}

For the Hamiltonian of Eq.~(\ref{eq:Hint}), 
the above procedure yields the following classical potential
\bsub
\label{Vpot}
\ba
\label{eq:V1}
V^{I}(\rho)/h_2 &=& 
\bz^2 \beta^2 - 
\rho\bz^2 \sqrt{2\!-\!\beta^2} \beta^3\Gamma
\nonumber\\
&& + \textstyle{\frac{1}{2}}(1\!-\!\bz^2)\beta^4~,\\
\label{eq:V2}
V^{II}(\xi)/h_2 &=&  
\bz^2[1 - \xi(1\!+\!\bz^2)] \beta^2 
-\bz \sqrt{2\!-\!\beta^2} \beta^3\Gamma
\qquad
\nonumber\\
&&
+ \textstyle{\frac{1}{4}}[2(1\!-\!\bz^2) + \xi(1\!+\!\bz^2)^2]\beta^4 
+ \xi\bz^4~,\quad
\ea
\esub
where $\Gamma\!\equiv\!\cos{3\gamma}$. 
The variables $\beta\in[0,\sqrt{2}],\,\gamma\in[0,2\pi)$ 
can be interpreted as polar coordinates 
in an abstract plane parametrized by Cartesian coordinates 
$x\!=\!\beta\cos{\gamma}$ and $y\!=\!\beta\sin{\gamma}$. 
The potential $V^{I}(\rho)$ [$V^{II}(\xi)$] 
has a global spherical [deformed] minimum with, respectively, 
$\beta_{\mathrm{eq}}\!=\!0$ 
[$\beta_{\mathrm{eq}}\!>\!0,\gamma_{\mathrm{eq}}\!=\!0$]. 
At the spinodal point ($\rho^{*}$), 
$V^{I}(\rho)$ develops an additional local deformed minimum, 
and the two minima become degenerate 
at the critical point $\rho_c$ (or~$\xi_c$). The spherical minimum 
turns local in $V^{II}(\xi)$ for $\xi>\xi_c$ 
and disappears at the anti-spinodal point~($\xi^{**}$). 
The  order parameter ${\beta_{\mathrm eq}}$, shown in Fig.~\ref{fig:1}, 
is a double-valued function 
in the coexistence region (in-between $\rho^{*}$ and $\xi^{**}$) 
and a step-function outside it. 
The potentials $V(\beta,\gamma=0)\!=\!V(x,y=0)$ 
for several values of $\xi,\rho$, are shown at the bottom rows 
of Figs.~\ref{fig:2}-\ref{fig:3} [panels (a)-(e)]. 
The height of the barrier at the critical point is 
$V_b\!=\!h_{2}[1-(1+\beta_0^2)^{1/2}]^2/2$. 
Henceforth, we set $\beta_0\!=\!1.35$ 
which is a typical value within the acceptable range, 
$1\leq \beta_0 \leq 1.41$, for deformed nuclei.
In this case, $V_b/h_2 \!=\!0.231$ 
(compared to $V_b/h_2 \!=\!0.0018$ in previous 
works~\cite{ref:Whel93}). 

The classical dynamics of $L\!=\!0$ vibrations, governed by 
$\hat{H}_{\mathrm{int}}$~(\ref{eq:Hint}), can be depicted conveniently 
via Poincar\'e sections~\cite{ref:Reic92}. 
These are shown for selected energies 
below the domain boundary $E_\mathrm{lim}\!=\!V(\beta\!=\!\sqrt{2},\gamma)$, 
and  control parameters, $\rho\!\leq\!\rho_c$ in Fig.~\ref{fig:2} 
and $\xi\!>\!\xi_c$ in Fig.~\ref{fig:3}. 
For $\rho\!=\!0$, the system is integrable, with 
$V^{I}(\rho=0)\!\propto\!\bz^2 \beta^2 + 
\textstyle{\frac{1}{2}}(1\!-\!\bz^2)\beta^4$. 
The sections for $\rho\!=\!0.03$ in Fig.~\ref{fig:1}, 
show the phase space 
portrait typical of an anharmonic (quartic) oscillator (AO) 
with two major regular islands, weakly perturbed by the 
small $\rho\cos 3\gamma$ term.
For small~$\beta$, 
$V^{I}(\rho)\!\approx\! \beta^2 \!-\! 
\rho\sqrt{2}\beta^3\cos 3\gamma$. 
The derived phase-space portrait, shown for $\rho\!=\!0.2$ 
in Fig.~\ref{fig:2},  
is similar to the  H\'enon-Heiles system (HH)~\cite{ref:Heno64} 
with regularity at low energy [panels (b$_1$)-(b$_2$)] and 
marked onset of chaos at higher energies [panels (b$_3$)-(b$_5$)]. 
The chaotic component of the dynamics increases with $\rho$ and 
maximizes at the spinodal point $\rho^{*}\!=\!0.546$. 
The dynamics changes profoundly in the coexistence region, 
shown for $\rho\!=\!0.65,\, 0.741$ in Fig.~\ref{fig:2}
and $\xi\!=\! 0.05$ in Fig.~\ref{fig:3}. 
As the local deformed minimum develops, robustly regular dynamics 
attached to it appears. The trajectories form a single island 
and remain regular at energies well above the barrier height~$V_b$, 
clearly separated from the surrounding chaotic environment. 
As $\xi$ increases, the spherical minimum becomes shallower, 
the HH-like dynamics diminishes and disappears 
at the anti-spinodal point $\xi^{**}\!=\!0.354$. 
Regular motion prevails for $\xi \!>\! \xi^{**}$, where the section 
landscape changes from a single to several regular islands. 
The dynamics is sensitive to local degeneracies of normal-modes, 
as can be seen by comparing Fig.~3($\mathrm{d_1}$) 
for $\xi\!=\!1.094$, corresponding to 
$\epsilon_{\beta}\!=\!\epsilon_{\gamma}$, 
with Fig.~3($\mathrm{e_1}$) for $\xi\!=\!1.1$. 

The quantum manifestations of the above rich classical dynamics 
can be studied via Peres lattices $\{x_i,E_i\}$~\cite{ref:Peres84}. 
Here $E_i$ are the energies of eigenstates~$\ket{i}$ of the Hamiltonian 
and $x_i \!\equiv\! \sqrt{2\bra{i}\hat{n}_d\ket{i}/N}$. 
The lattices can distinguish regular from irregular states 
by means of ordered patterns and disordered 
meshes of points, respectively~\cite{ref:Peres84,ref:Stran09}. 
The particular choice of $x_i$ can associate the states with 
a given region in phase space through the classical-quantum correspondence 
$\beta \!=\!x \!\leftrightarrow\! x_i$, obtained by 
comparing with the expectation 
value of $\hat{n}_d$ in the static condensate~\cite{ref:MacLev11}.
The Peres lattices for $L\!=\!0$ eigenstates of 
$\hat{H}_\mathrm{int}$~(\ref{eq:Hint}) 
with $N\!=\!80$, 
are shown on the bottom rows of Figs.~\ref{fig:2}-\ref{fig:3}, 
overlayed on the classical potentials $V(x,y=0)$.
For $\rho\!=\!0$, the Hamiltonian~(\ref{eq:H1}) 
has U(5) dynamical symmetry with a solvable spectrum 
$E_i \!=\! 2\bar{h}_2[\bz^2N\!-\!1 + (1\!-\!\bz^2)n_d]n_d$. 
For large $N$ and replacing $x_i$ by $\beta$, the Peres 
lattice coincides with $V^{I}(\rho=0)$, a trend seen in Fig.~2(a). 
Whenever a deformed minimum occurs in the potential, the Peres lattices 
exhibit regular sequences of states, localized in the region of 
the deformed well and persisting 
to energies well above the barrier. 
They are related to the regular islands in the Poincar\'e sections 
and are well separated from the remaining states, which form 
disordered (chaotic) meshes of points at high energy. 
The number of such sequences 
is larger when the potential well is deeper. 
The regular $L\!=\!0$ states form 
bandheads of rotational sequences $L\!=\!0,2,4,\ldots$ 
($K\!=\!0$ bands).
Additional $K$-bands with $L\!=\!K,K+1,K+2,\ldots$ 
can also be identified. 
An example of such regular $K\!=\!0,2$ bands for 
$\hat{H}_{\mathrm{int}}$ at the critical point, is shown in Fig.~4(a).
In the nuclear physics 
terminology, the lowest $K\!=\!0$ band refers to the ground band and 
excited $K$-bands correspond to multiple 
$\beta$ and $\gamma$ vibrations 
about the deformed shape with angular momentum 
projection $K$ along the symmetry axis. 
The states in each band share a common intrinsic 
structure as indicated by their nearly equal values of 
$\langle \hat{n}_d \rangle$, and a coherent decomposition of their wave 
functions in the rotor basis~\cite{ref:MacLev12}. 
The occurrence of such a pure ordered~band structure amidst a complicated 
environment, indicates the relevance, for QPTs,  
of an adiabatic separation of modes~\cite{ref:Mac10} 
and possibly partial symmetries~\cite{ref:Lev07}, 
for a subset of states.
\begin{figure*}[!t]
\begin{center}
\includegraphics[width=\linewidth]{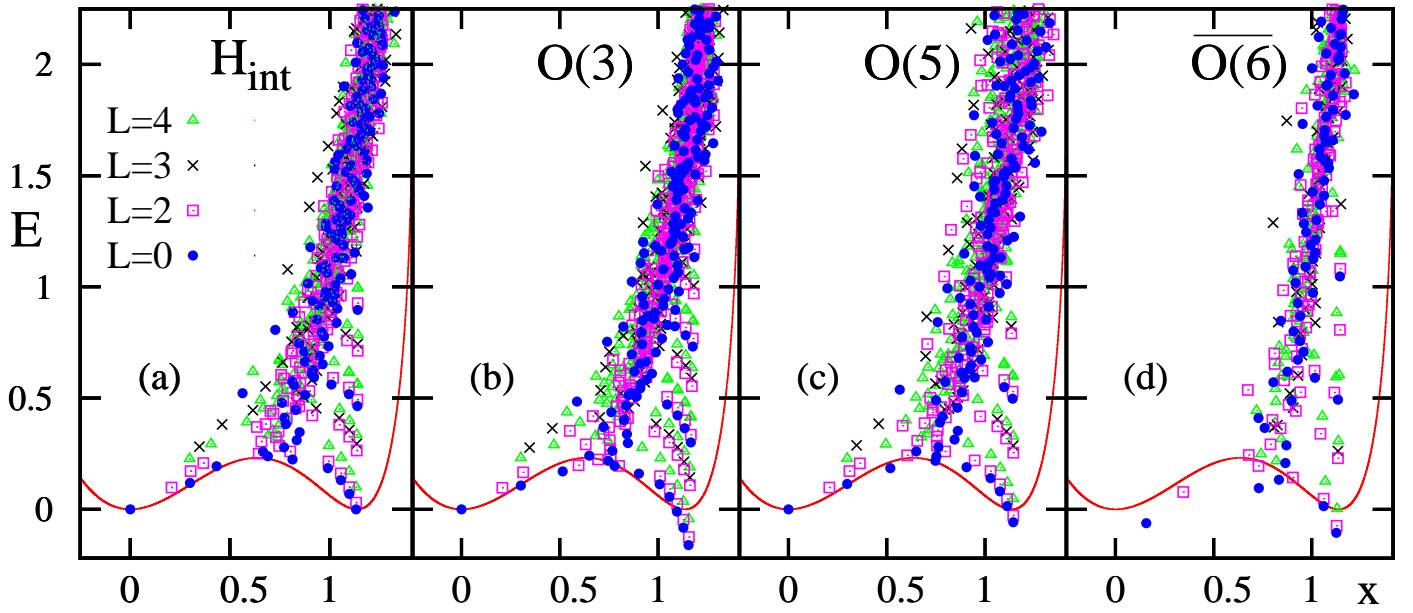}
\end{center}
\caption{(Color online). 
Peres lattices $\{x_i,E_i\}$ for $N\!=\!50,\, L\!=\!0,2,3,4$ 
eigenstates of 
$\hat{H}_\mathrm{int}^{I}(\rho=\rho_c) \!=\! 
\hat{H}_\mathrm{int}^{II}(\xi=\xi_c)$ (\ref{eq:Hint}), 
with $h_2\!=\!1,\,\beta_0\!=\!1.35$ [panel (a)] and additional 
collective terms, Eq.~(\ref{eq:Hcol}), with $c_i/h_{2}\!=\!1$, 
involving $\mathrm{O(3)},\,\mathrm{O(5)}$ and 
$\overline{\mathrm{O(6)}}$ rotations 
[panels (b), (c), and~(d)]. 
To enhance visibility, a small 
energy shift $\Delta E_L \!=\! 0.005 L(L+1)$ is added. 
The classical potential shown, is the same in all cases. 
Notice in panels (a)-(b)-(c), 
the well-developed rotational bands 
($K\!=\!0$, $L\!=\!0,2,4$) 
and ($K\!=\!2$, $L\!=\!2,3,4$) formed by the regular states 
in the deformed phase, which are distorted 
in panel~(d).}
\label{fig:4}
\end{figure*}

So far the discussion involved the intrinsic part of the Hamiltonian 
(\ref{eq:H}). The collective part, which does not affect $V(\beta,\gamma)$ 
can be transcribed in the form~\cite{ref:Lev87}
\ba
\hat{H}_\mathrm{col} &=& \bar{c}_3 
[\,\hat{C}_{\mathrm{O(3)}} - 6\hat{n}_d \,] + 
\bar{c}_5[\,\hat{C}_{\mathrm{O(5)}} - 4\hat{n}_d\, ] 
\nonumber\\
&&
+\, \bar{c}_6 [\,\hat{C}_{\overline{\mathrm{O(6)}}} - 5\hat{N}\,] ~, 
\label{eq:Hcol}
\ea
where $\hat{C}_{G}$ denotes the quadratic Casimir operator of the 
group $G$, as defined in~\cite{ref:Lev87} and 
$\bar{c}_i\equiv c_{i}/N(N-1)$. 
The kinetic $\mathrm{O(3),\,O(5)}$ and $\overline{\mathrm{O(6)}}$ 
terms involve collective rotations associated with the Euler angles, 
$\gamma$ and $\beta$ degrees of freedom, respectively. 
Fig.~\ref{fig:4} shows the Peres lattices corresponding to 
$L\!=\! 0,2,3,4$ eigenstates of 
$\hat{H}_{\mathrm{int}}$ at the critical-point, plus added 
rotational terms one at a time. As seen in Figs.~4(b)-4(c), the 
$c_3$ and $c_5$ terms preserve the, previously 
mentioned, ordered $K$-bands of $\hat{H}_{\mathrm{int}}$, Fig.~4(a). 
The calculated spectrum of these bands resembles a rigid rotor 
[$L(L+1)$ splitting] for the 
$\mathrm{O(3)}$ term and a rotor with centrifugal stretching for the 
$\mathrm{O(5)}$ term~\cite{ref:Lev06}. 
In contrast, the regular band-structure is strongly disrupted 
by the $\overline{\mathrm{O(6)}}$ term [Fig.~4(d)]. The latter 
couples the deformed  and spherical configurations~\cite{ref:Lev06} 
and mixes strongly the regular and irregular states. 
Only the $\overline{\mathrm{O(6)}}$ rotations involve the motion in 
the $\beta$ variable~\cite{ref:Lev87}, 
highlighting the importance, in QPTs, of the coupling of the order 
parameter fluctuations with soft modes~\cite{ref:Belitz05}.
These results demonstrate the advantage of using the 
resolution of the Hamiltonian (\ref{eq:H}) in studies of QPTs, since 
a strong $\overline{\mathrm{O(6)}}$ term in the collective part can 
obscure the simple patterns of the dynamics disclosed by the 
intrinsic part. 

In summary, we have presented a comprehensive analysis of the 
dynamics across a generic first order QPT between stable spherical and 
deformed configurations in the IBM framework. 
The intrinsic part of the Hamiltonian determines the Landau potential, 
and its classical analysis reveals a change in the system from an 
AO- and HH-type of dynamics on the 
spherical side, into a pronounced regular dynamics on the deformed side 
of the transition. The dynamics inside the coexistence region 
is robustly regular and confined to the deformed well, 
in marked separation from the chaotic behavior ascribed to the 
spherical well. The coexistence of regular and chaotic motion persists 
in a broad energy range throughout the coexistence region and is absent 
outside it. This simple pattern manifests itself also in the quantum 
analysis, disclosing regular rotational bands in the deformed region, 
which persist to energies well above the barrier and retain their 
identity amidst a complicated environment.
These ramifications of a divided phase space structure are observed at 
any $\beta_0>0$, but are more pronounced for higher barriers 
(larger $\beta_0$)~\cite{ref:MacLev11,ref:MacLev12}. 
Kinetic terms in the collective part of the Hamiltonian involving 
rotations in the orientation (Euler angles) and triaxiality ($\gamma$) 
variables, preserve the ordered band-structure, while collective 
rotations in the deformation ($\beta$) variable can disrupt it 
by mixing regular and irregular states. 
Our results are of interest not only for nuclei, but also to other 
interacting systems undergoing a first order QPT. They demonstrate 
a clear connection between order, chaos and structural changes 
of coexisting phases in such systems.   

This work is supported by the Israel Science Foundation. 
M.M. acknowledges the Golda Meir Fellowship Fund and partial support 
by the Czech Ministry of Education (MSM 0021620859).

\end{document}